\documentstyle [prl,aps,floats,twocolumn]{revtex}
\input psfig.tex
\begin{document}

\def\ba{\begin{eqnarray}}
\def\ea{\end{eqnarray}}
\def\be{\begin{equation}}
\def\ee{\end{equation}}
\def\({\left(}
\def\){\right)}
\def\[{\left[}
\def\]{\right]}
\def\lagrange {{\cal L}}
\def\del {\nabla}
\def\d {\partial}
\def\Tr{{\rm Tr}}
\def\half{{1\over 2}}
\def\fourth{{1\over 8}}
\def\bibi{\bibitem}
\def\S{{\cal S}}
\def\xx{\mbox{\boldmath $x$}}
\newcommand{\labeq}[1] {\label{eq:#1}}
\newcommand{\eqn}[1] {(\ref{eq:#1})}
\newcommand{\labfig}[1] {\label{fig:#1}}
\newcommand{\fig}[1] {\ref{fig:#1}}
\newcommand\bigdot[1] {\stackrel{\mbox{{\huge .}}}{#1}}
\newcommand\bigddot[1] {\stackrel{\mbox{{\huge ..}}}{#1}}

\twocolumn[\hsize\textwidth\columnwidth\hsize\csname @twocolumnfalse\endcsname
\preprint{DAMTP-97-23, SWAT-97-152, arch-ive/9704012}
\title{Normal Modes of the B=4 Skyrme Soliton} \author{Chris
Barnes\thanks{email:barnes\@puhep1.princeton.edu}, Kim
Baskerville\thanks{email: phkim\@python.swan.ac.uk}, and Neil
Turok\thanks{email:N.G.Turok\@amtp.cam.ac.uk}} \address{${}^*$Joseph Henry
Laboratory, Princeton University, Princeton, NJ, USA 08540\\
 ${}^\dagger$Physics Department, University of Wales Swansea, Singleton Park,
Swansea SA2 8PP\\ 
${}^\ddagger$DAMTP, Silver St,Cambridge, CB3 9EW, U.K.  }
\date\today 
\maketitle

\begin{abstract}
The Skyrme model of nuclear physics requires quantisation 
if it is to match observed nuclear properties. 
A simple technique 
is used to 
find the normal mode spectrum of the baryon number $B=4$ Skyrme
soliton,
representing the $\alpha$ particle. 
We find sixteen vibrational modes and classify them under the cubic
symmetry group $O_h$ of the static solution. The spectrum 
possesses a remarkable structure, with 
the lowest energy modes lying in those representations
expected from an approximate correspondence between
Skyrmions and BPS monopoles. The next mode up is the `breather', 
and above that are
higher multipole breathing modes.
\end{abstract}
\vskip .2in
]

\section{Introduction}
In the Skyrme model of nuclear physics, both pions and nucleons are
represented by a single scalar SU(2) group valued field, $U(x)$.
Pions occur as field quanta, while baryons are instead represented as
topological solitons.  The classical Skyrme theory, with a simple
quantisation of the spin and isospin collective modes, provides a
description of nucleons and the $\Delta$ resonance in modest agreement
with experiment \cite{ANW,AN}.

Applying the Skyrme model to larger nuclei and to nuclear matter is an
even more interesting proposition, since with no additional free
parameters one could compare the theory with the binding
energies and gamma ray spectra of all nuclei.  There has been
progress in understanding the structure of Skyrme multisolitons
\cite{Battye}, \cite{Braaten}, but it is clear that unless
quantum fluctuations about the static solutions are included, there is
little chance of success.  For real nuclei the relative kinetic
energies of the nucleons in the ground state are large, so a
quantisation at least of a number of degrees of freedom equal to the
number of nucleon coordinates is essential. Note that the simple 
collective coordinate quantisations of spin and isospin  \cite{ANW,AN}
include effects of order $\hbar^2$, while ignoring
effects of order $\hbar$.

Recently there has been some progress in quantisation, based upon
Manton's notion of representing low energy solitonic excitations as
motion on a finite dimensional space of moduli.  A study of
the deuteron by Leese et. al. \cite{Leese} using an instanton
approximation for the field configurations, gave encouraging agreement
with experimental properties, but only included two of the four
expected vibrational modes of the deuteron.  Walet \cite{Walet}
extended this treatment to estimate all vibrational frequencies for
the deuteron and triton, again using the instanton approximation.

In this Letter we follow a different track.  We directly compute the
low energy normal modes, finding their frequencies and the
representations they lie in under the static soliton's symmetry group.
The frequencies and representations provide a coordinate-independent
description of the configuration space around the static solution; in
the harmonic approximation they determine the quantum vibrational
energy levels.  Our results provide new insight to the moduli space
approach, since the representations for the lowest frequency modes
turn out to be just those expected from a recently understood 
approximate  correspondence between
Skyrmions and BPS monopoles \cite{Manpri}. Furthermore, the success of
these calculations allows one to contemplate going beyond the moduli
space approximation and performing a full semiclassical quantisation
of the field theory. This is an attractive goal, since the Skyrme theory
could then be incorporated in the framework of chiral effective
Lagrangians \cite{Gasser}, allowing a unified treatment of mesons,
baryons and higher nuclei.

\section{Method}
Since the SU(2) manifold is a 3-sphere, we represent it 
in terms of a scalar field $\phi\
\in\ {\bf R}^4$, with $\phi^a\phi^a = 1$.  In terms of this field, the
Skyrme Lagrangian density is
\ba
&\lagrange = {1\over 2} \d_\mu\phi\cdot\d^\mu\phi+  
 \omega^2_\pi\phi^1 + \lambda(\phi\cdot \phi - 1) + {}
&\nonumber\\
& {1\over 4}\bigl\{\(\d_\mu\phi\cdot \d_\nu\phi\)
\(\d^\mu\phi\cdot \d^\nu\phi\)  
	- \(\d_\mu\phi\cdot\d^\mu\phi\)\(\d_\nu\phi\cdot\d^\nu\phi\)\bigr\}&
\labeq{lagrangian}\ea
with $\lambda$ a Lagrange multiplier field.
Here, length and time are in units 
of ${2\over e F_\pi}$, energy in units of ${F_\pi\over 2 e}$.
In these rescaled units, the only
remaining parameter is $\omega_\pi = {2 m_\pi \over e F_\pi}$, the
oscillation frequency for the homogenous pion field.
For the most part we have adopted  the `standard'
value \cite{AN} of $\omega_\pi =
.526$, although we have also performed calculations at 
twice this value.

The mixed space and time 
derivative terms in the Skyrme Lagrangian make 
numerical solution in general difficult. Isolating the time derivative terms
one has
\be
\lagrange = {1\over 2}\bigdot{\phi^a} K^{ab}(\d_i\phi)\bigdot{\phi^b} -
V(\phi,\d_i\phi)
+ \lambda(\phi \cdot \phi - 1)\ ,
\labeq{condensed}
\ee
where  $K^{ab}= \delta^{ab} (1+(\d_i\phi)^2) - \d_i\phi^a \d_i\phi^b$
is a local inertia matrix, and
$V(\phi,\d_i\phi)$ is the potential. In general, $K^{ab}$ is time dependent, 
but for small perturbations around a static solution $\phi_{\rm st}({\xx})$
we write 
$\phi(\xx,t) =
 \phi_{\rm st}(\xx) + \epsilon (\xx,t)$, $\epsilon \ll 1$ 
and to second order
in $\epsilon$ the Lagrangian is
\ba 
\lagrange &=& {1 \over 2} \bigdot{\epsilon^a}K^{ab}\(\d\phi_{\rm st}\)
  \bigdot{\epsilon^b} - V(\phi,\d_i\phi)
+ \lambda(\phi \cdot \phi - 1)
\labeq{continuousapprox}
\ea
This Lagrangian leads to classical field equations:
\be
K^{ab}(\d\phi_{\rm st})\bigddot{\phi^a} = \d_i\(\d V\over \d \phi^b_{\ ,i}\) 
- {\d V \over \d \phi^b}  + \lambda\phi^b  \ .
\ee
where the matrix $K^{ab}({\xx})$ is
taken to be its value at the static classical solution.
Equation (4) closely approximates the Skyrme equations for fields
near a static classical solution, precisely the desired regime for
studying soliton normal modes. 

In order to numerically solve the field equations, we discretise the
action, \eqn{continuousapprox}, using a diagonal differencing scheme
for the four spatial derivative terms, achieving a high degree of
locality and second order accuracy in both spatial and time steps. The
numerical code conserves energy and baryon number to within 1 part in
$10^5$ over the course of extremely long (50k timestep) runs.
Periodic boundary conditions are used.
We first create the appropriate minimal
energy static solution by straightforward time evolution from
four-Skyrmion initial conditions. 
A simple relaxation procedure sets the field momenta
zero each time the kinetic energy reaches a maximum. We find
the fields 
rapidly converge on the minimum energy configuration.
$K^{ab}$ is set equal to $\delta^{ab}$ in this part of the calculation,
since this does not affect the final static solution.

Next we slightly perturb the fields and 
evolve them forward again, but now using the full inertia 
matrix $K^{ab}(\d\phi_{\rm st})$. The evolving field is 
\be
\phi(\mbox{\boldmath $x$},t) = \phi_{\rm st}(\mbox{{\boldmath $x$}}) + 
\sum_{\rm modes} \epsilon_n\delta_n(\mbox{\boldmath $x$}){\rm cos} (\omega_n t)
+ {\bf O}(\epsilon^2) \ 
\labeq{timebehaviour}
\ee
where the functions 
$\delta_n(\mbox{\boldmath $x$})\in {\bf R}^4$, obeying 
 $\delta_n(\xx)\cdot\phi_{\rm st}(\xx)=0$, are the normal
modes, each excited with amplitude $\epsilon_n$.  The normal mode frequencies 
$\omega_n$ are found by Fourier transforming $\phi(\mbox{\boldmath $x$},t)$
with respect to time
at any point in the box, and plotting the resulting power spectrum.

The space of perturbations has a useful inner product
\be
\langle\delta_1 |\delta_2\rangle = \int_{\rm box} \delta_1^a(\xx) K^{ab}(\xx)\delta_2^b(\xx)\ {\rm d}^3 \xx,
\labeq{innerproduct}\ee
which is zero for normal modes $\delta_1$ and $\delta_2$ 
if $\omega_1 \neq \omega_2$.  The inner product allows one to determine
the degeneracies of the normal mode frequencies and 
the representations in which the modes transform under the soliton's 
symmetry group.

\section{Results for the B=4 soliton}\vskip-.4em

We have applied this technique to the case of the Skyrme soliton with
baryon number four: the $\alpha$ particle.  The $\alpha$ particle
provides an especially simple case for quantisation, because the
ground state possesses zero angular momentum and isospin. The static
soliton has cubic symmetry; its energy and baryon number density
concentrate along the edges of a cube
\cite{Braaten}.  The full 48
dimensional cubic group of symmetries $O_h$ (for notation see
\cite{Hamermesh}) is generated by 90 degree and 120
degree rotations, and parity $I$. After 
an appropriate global isospin rotation, the action 
of these group elements on spatial coordinates and
pion fields ($\phi^a = (\sigma, \vec{\pi})$) is
as follows \cite{LeeseMan}:
\ba
C_4: \quad (\pi^1,\pi^2,\pi^3)(x,y,z) &\rightarrow& (-\pi^2,-\pi^1,-\pi^3)(-y,x,z)\nonumber\\
C_3: \quad (\pi^1,\pi^2,\pi^3)(x,y,z) &\rightarrow& (\pi^2,\pi^3,\pi^1)(y,z,x)\nonumber\\
I: \quad (\pi^1,\pi^2,\pi^3)(x,y,z) &\rightarrow& (\tilde{\pi}^1,\tilde{\pi}^2
,\tilde{\pi}^3)(-x,-y,-z)\nonumber
\ea
with $\tilde{\pi}^1={1\over 3} (\pi^1-2\pi^2-2\pi^3)$ etc.
From this it is straightforward to check that a homogeneous pion field 
falls into the two dimensional representation $E^+$ and the 
one dimensional representation $A_2^-$, where superscripts indicate
parity.

\begin{figure}
\centerline{\psfig{file=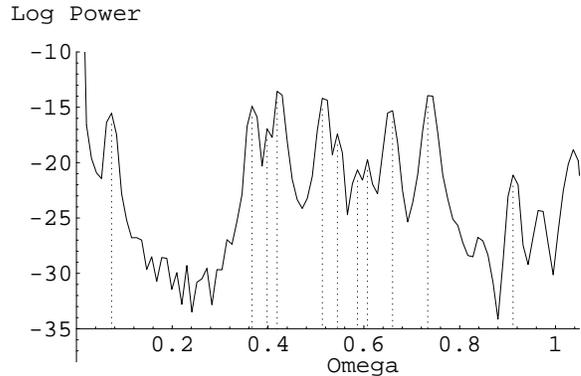,width=3.in}}
\caption{Fourier power spectrum for perturbations around a 
B=4 soliton.  Frequency
is in Skyrme units, power scale is arbitrary. Note that the frequency of
homogeneous pion oscillations is $\omega_\pi = .526$ in 
these units.}
\labfig{spectrum}
\end{figure}

A typical power spectrum of the perturbations is shown in Figure
\fig{spectrum}. Spectra at different sites and for different field
components show the same peaks, but with differing heights.  Once the
normal mode frequencies $\omega_n$ are identified, maps of the normal
modes $\delta_n(\xx)$ may be constructed by performing discrete
Fourier sums on each component of the field as it evolves.  For
degenerate modes, each set of perturbed initial conditions gives a
different linear combination $\sum_i \epsilon_i \delta_i(\xx)$ of
modes with the same frequency. Other linear
combinations 
may also be produced by applying the symmetries of the static soliton $\S_1,
\S_2, ...$ to a given mode $\delta$.  The degeneracy of a given
frequency is found by computing the rank of the matrix of inner products
between the different linear combinations of degenerate modes produced
in these ways.  Once the degeneracy is determined, a complete
orthonormal basis of modes with this frequency can be constructed.
The character of any $O_h$ group element can then be
computed as a trace. These characters were  within $\pm .001$
of an integer value, and interpretation was unambiguous.

Each peak in Figure \fig{spectrum} marks the frequency of a normal
mode.  The lowest peak (at $\omega=.07$) is the rotational zero mode,
shifted to nonzero frequency by finite size effects, effectively through
interactions with image solitons one box length away.
There are also
several peaks corresponding to relatively delocalised modes which we
interpret as pion radiation. 
Two (at $\omega =
.545,.587$) correspond to the lowest radiation modes, homogeneous 
away from the soliton, whose threefold
degeneracy is split
into $E^++A_2^-$ by the presence of the soliton. 
The first radiation mode at nonzero wavenumber is
at $\omega =.908$. 
The remaining modes are the true
vibrational excitations of the $\alpha$ particle.  Somewhat
fortuitously, the box size was small enough that the lowest
inhomogeneous radiation mode has a frequency above the highest
vibrational mode.

Four widely separated $B=1$ Skrymions have 24 zero modes, corresponding
to 3 translations and 3 isorotations each. 
As they combine to form the
B=4 soliton, 9 of them (3 translations, 3 rotations and 3
isorotations) remain as zero modes of the new system. 
One might expect the remaining 15 modes would survive as
low energy vibrational modes. This is one fewer than what we find: we
have an additional breather mode.

The vibrational modes distort the $B=4$ soliton as
illustrated in Figure \fig{bigplot}, and explained in the Table. The
modes naturally divide into two sets.  The lower nine vibrational
modes consist of deformations which, roughly, involve
incompressible flow of the baryon charge. In contrast, 
the higher seven vibrational modes
all have a `breathing' character, in which local baryon charge expands
or contracts to occupy a greater or lesser volume.  The breather
itself is simply a rescaling of the size of the soliton, with
consequent change in density. The next mode up involves breathing
motion of a dipole character, and the one above that of a
quadrupole nature.

Remarkably, the vibrational
modes below the breather
fall into representations corresponding to those for
small zero-mode deformations of the BPS 4-monopole solution
\cite{Manpri}. The same 
phenomenon occurs in the deuteron
\cite{deutpaper}, and it will be straightforward to check
the $B=3,5,6,7$ solitons using the  methods described here.
Qualitative similarities between
Skyrme multisolitons and BPS multimonopoles
and their scattering dynamics have been
noted before \cite{Battye1},\cite{Battye}. Our finding suggests
a connection between the lowest energy Skyrmion vibrational modes
and the multimonopole moduli spaces.
If it holds
for higher nuclei, there will be $4B-7$ such modes.
It would be very interesting to interpret this
number in terms of individual nucleon degrees of freedom (presumably
translations and spin/isospin).

We have investigated the effect of doubling the parameter $\omega_\pi$
on the spectrum of modes. All modes move up in frequency. The
nine lowest modes move up by 15-25 per cent, the breathing modes 
by 30-45 per cent, 
and the homogeneous pion modes roughly double in frequency. 
This has important phenomenological 
consequences.
The lowest lying vibrational level for the
real $\alpha$ particle is at 20.1 MeV, whereas for $\omega_\pi = .526$
our lowest lying frequency is at $0.7 \omega_\pi$, or 94 MeV. For
$\omega_\pi= 1.052$ this is improved to 63 MeV.  Whilst finite volume
and grid spacing effects are present at a level of a few per cent in
these results, we can safely conclude that a fit between the Skyrme
model and nuclear gamma ray spectra will require parameters different
than those usually used.  It is interesting that a study of a single
Skyrmion's breather mode, attempting to identify it with the `Roper
resonance', also found that large values of $\omega_\pi \sim 1.5$ were
required \cite{Breit}.  It remains to be seen whether such large
values of $\omega_\pi$ can be accomodated in the theory.

We thank Richard Battye, Guy Moore, Conor Houghton, Paul Sutcliffe
and especially Nick Manton for helpful discussions. We also
acknowledge the Pittsburgh Supercomputing Center grant \#AST9G3P for
CRAY T3D supercomputer time.

\newcommand\sfig[1]{\psfig{file=#1,width=1.135in}}
\begin{figure*}[t]
\begin{tabular}{ccccc}
  Static $B=4$ soliton, $\phi_{\rm st}$ &  & 
$\phi_{\rm st}$ $+$, $-$ $\delta_{.367}^1$ & & 
$\phi_{\rm st}$ $+$, $-$ $\delta_{.367}^2$ 
\\
\sfig{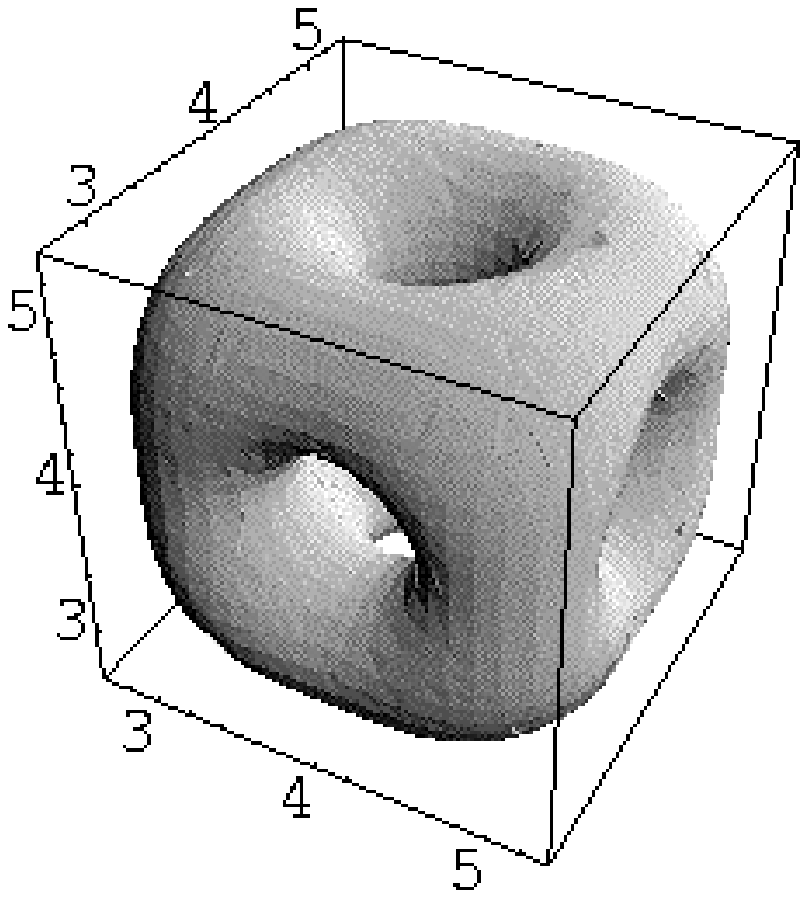} & &
\hbox{\sfig{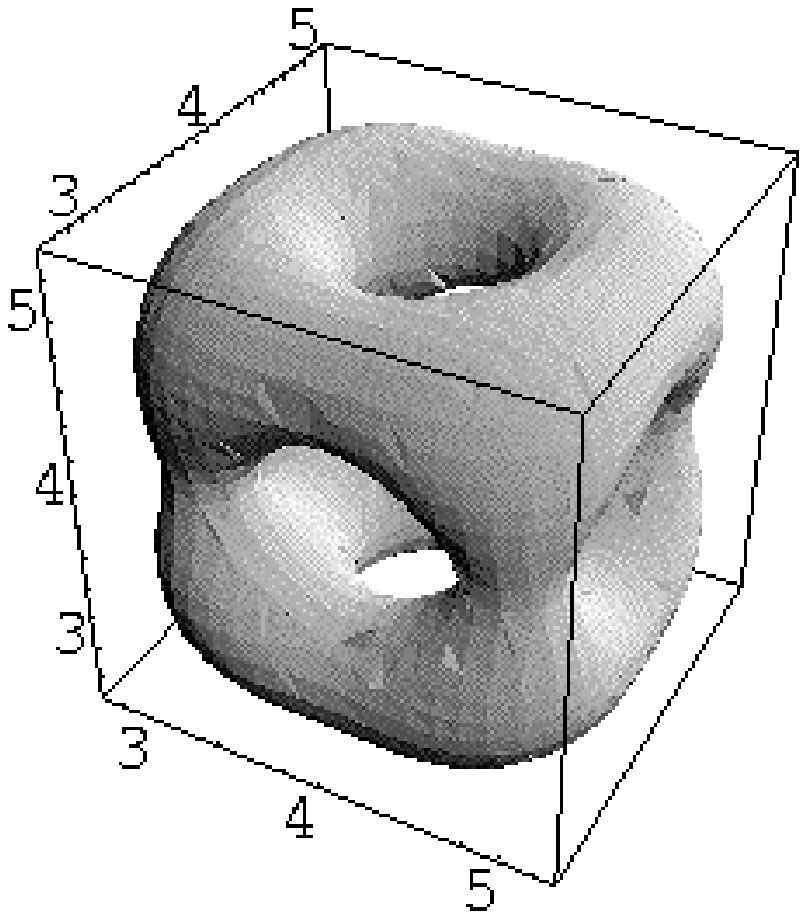}\hfil\sfig{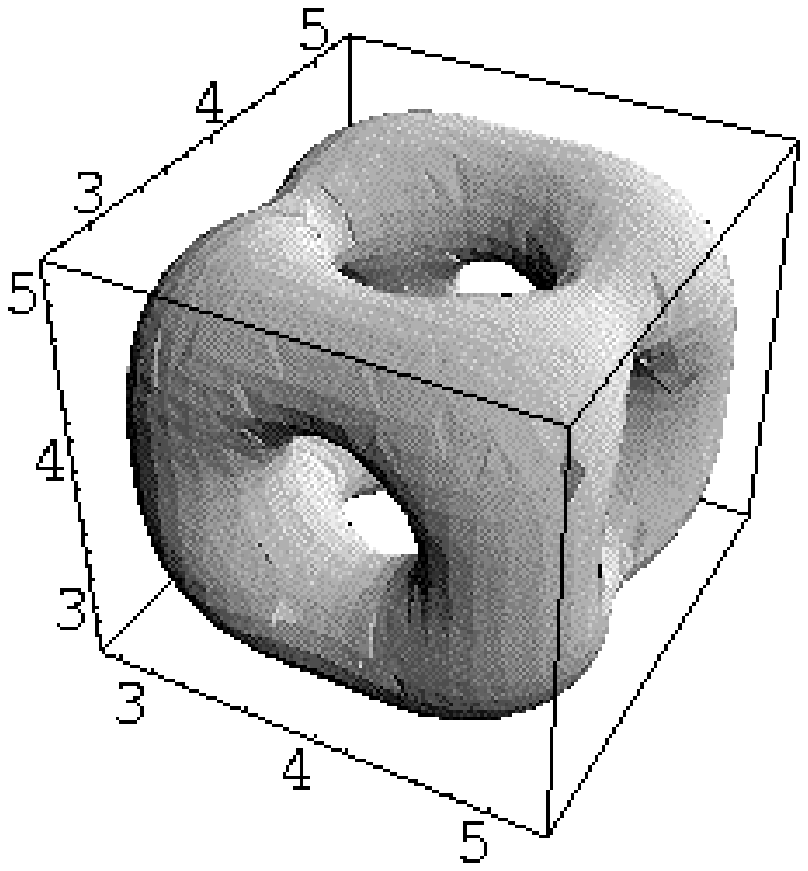}} & &
\hbox{\sfig{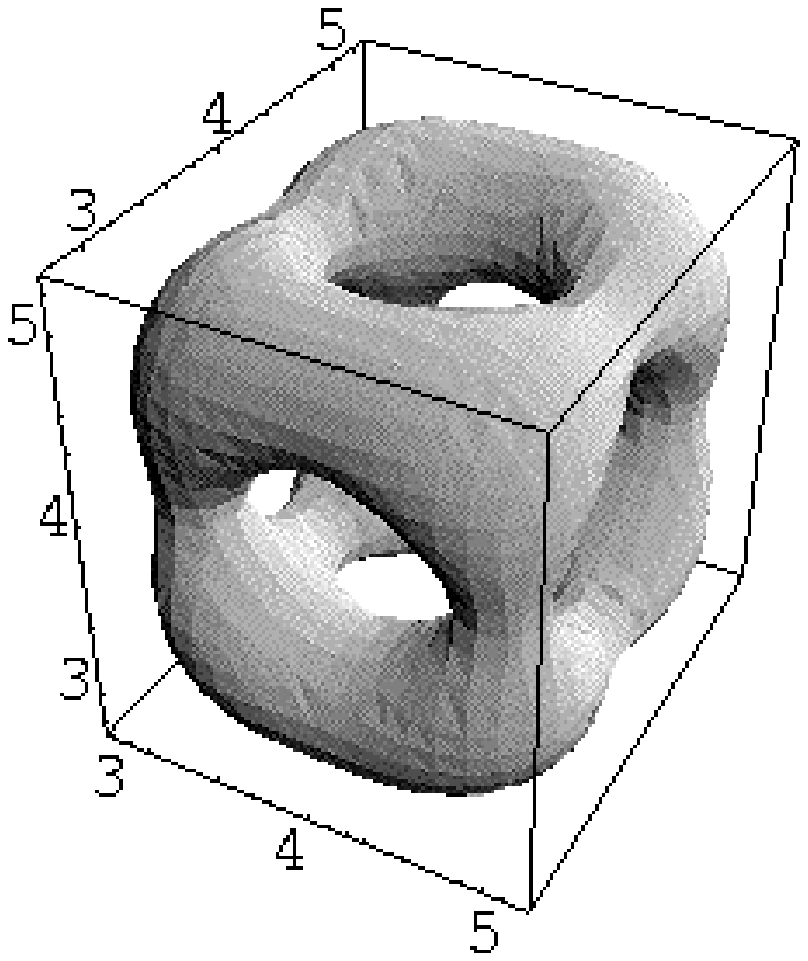}\hfil\sfig{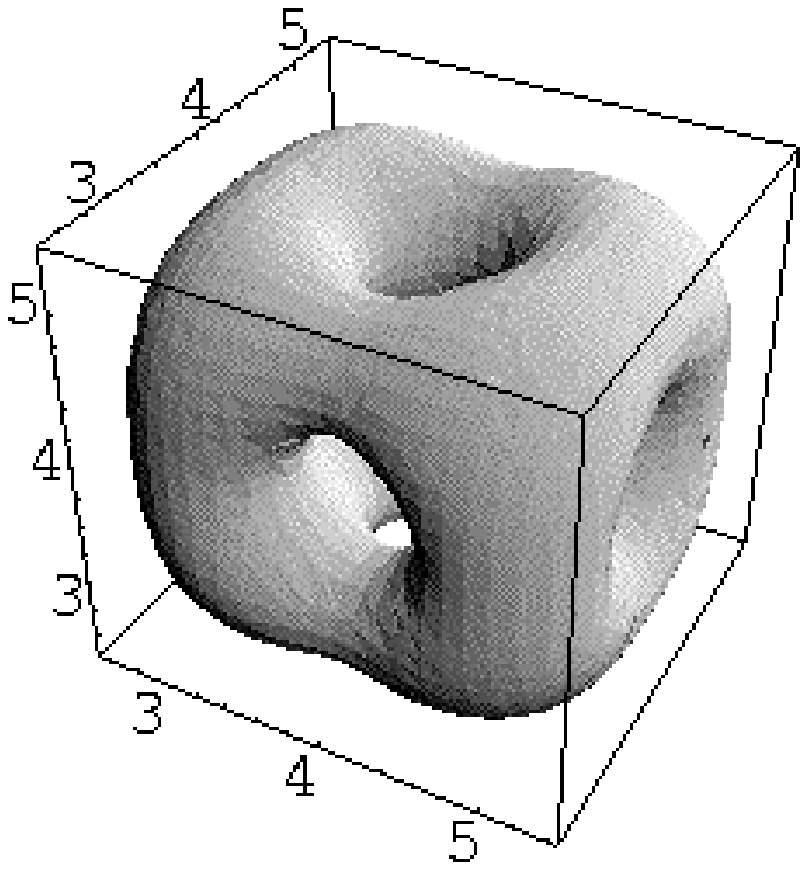}} \\
 & \\   
$\phi_{\rm st}$ $+$, $-$ $\delta_{.405}$  & &
  $\phi_{\rm st}$ $+$, $-$ $\delta_{.419}$ & &
 $\phi_{\rm st}$ $+$, $-$ $\delta_{.513}$ 
\\
\hbox{\sfig{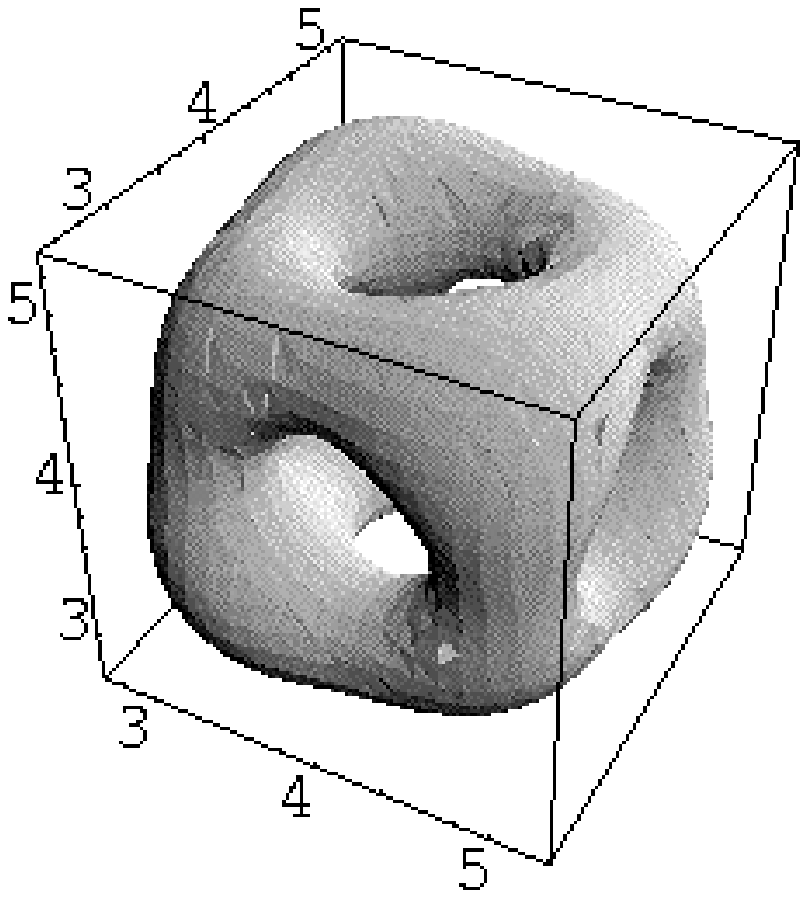}\hfil\sfig{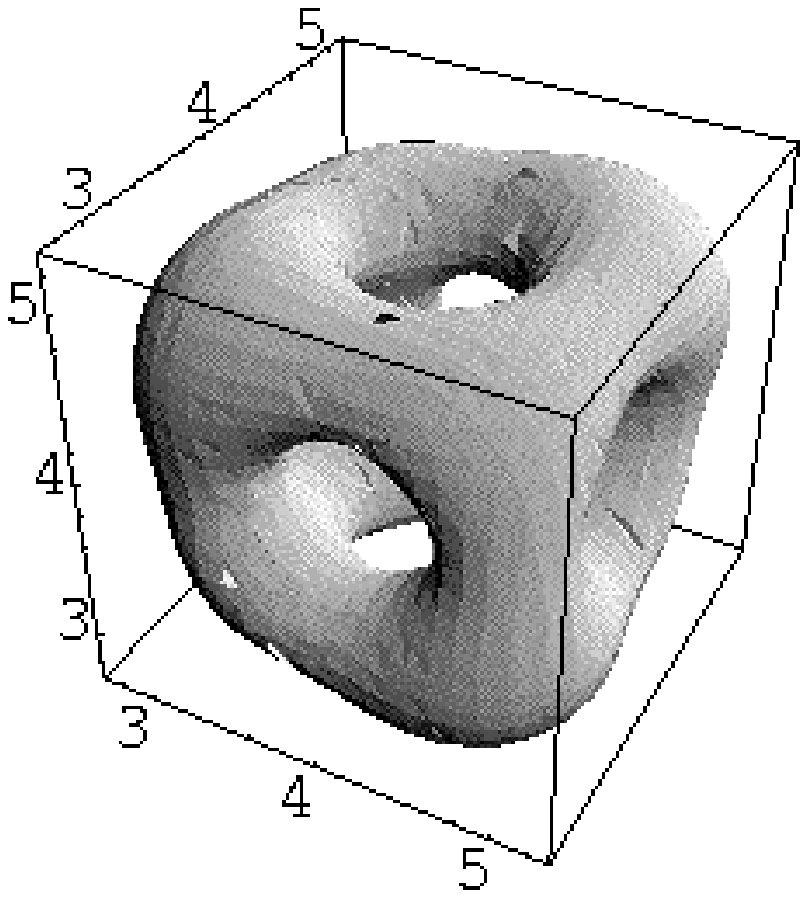}} & &
\hbox{\sfig{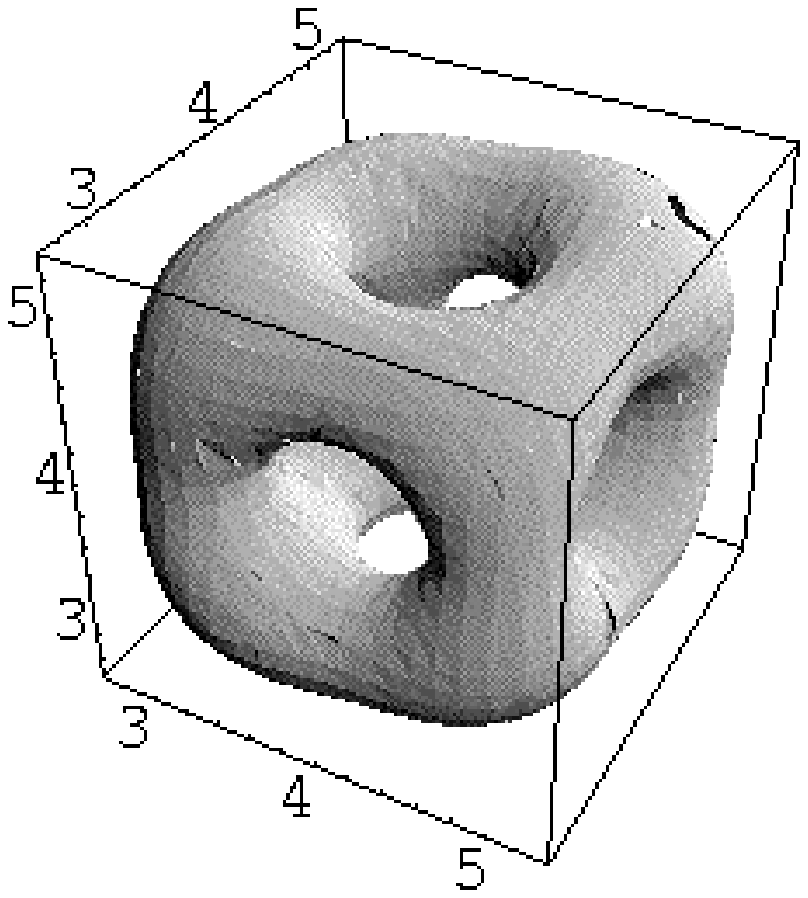}\hfil\sfig{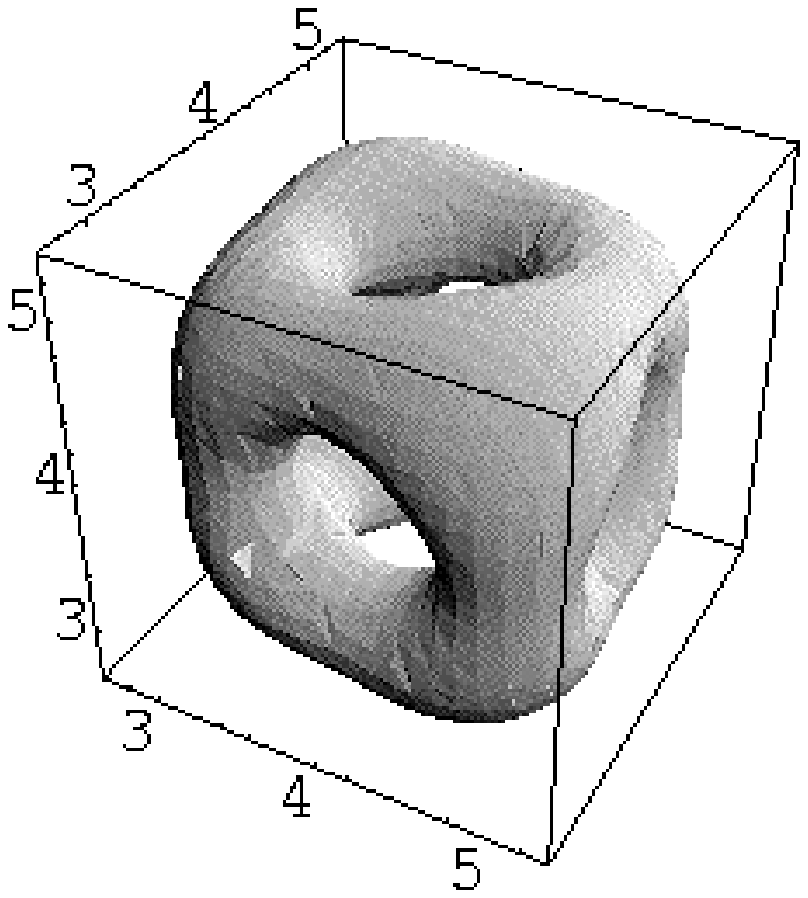}} & &
\hbox{\sfig{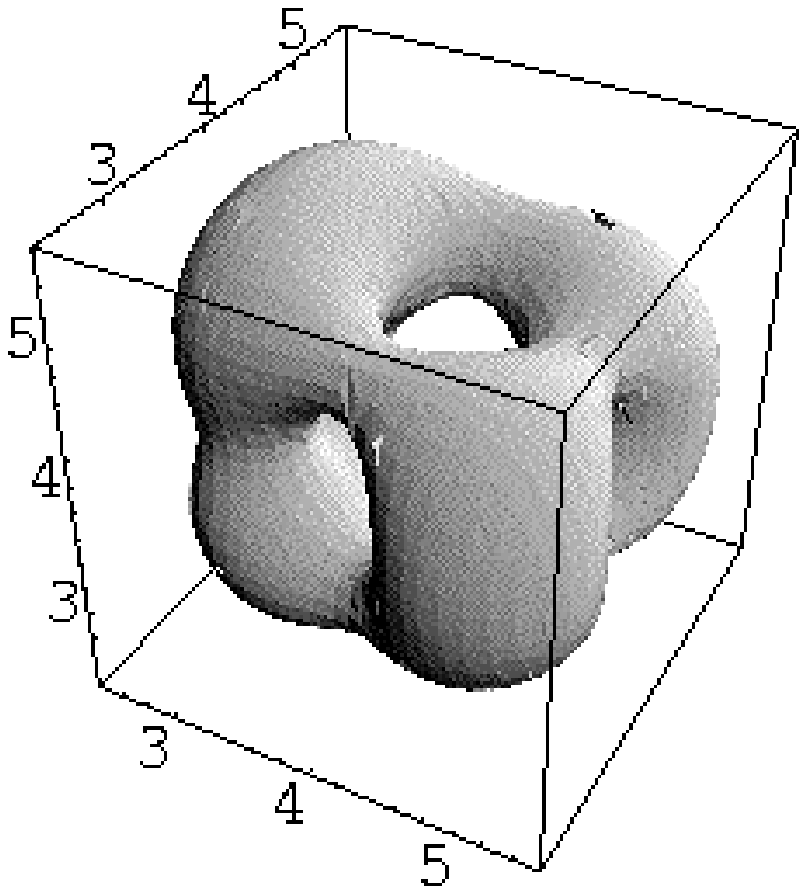}\hfil\sfig{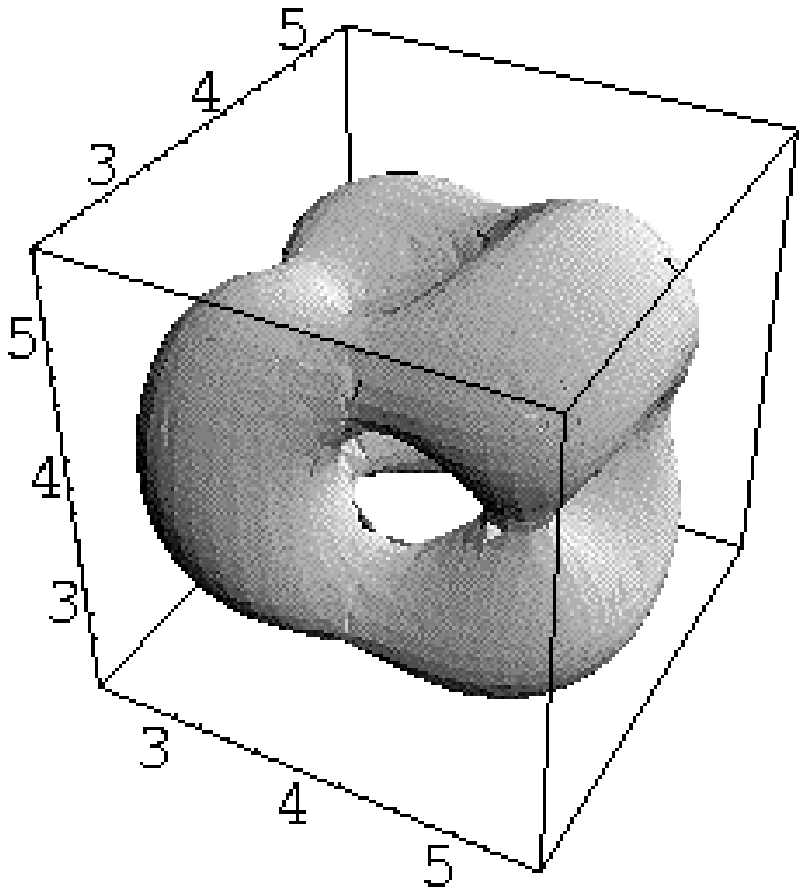}} \\
 & \\   
$\phi_{\rm st}$ $+$, $-$ $\delta_{.605}$ 
& & $\phi_{\rm st}$ $+$, $-$ $\delta_{.655}$ & & $\phi_{\rm st}$ $+$, $-$ $\delta_{.738}$ \\
\hbox{\sfig{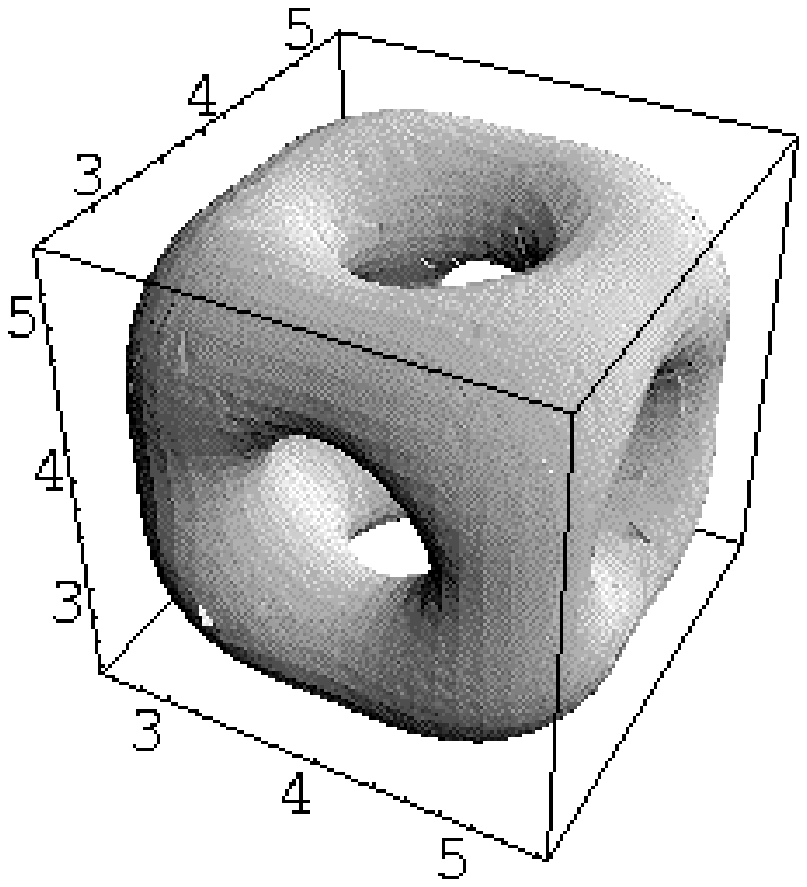}\hfil\sfig{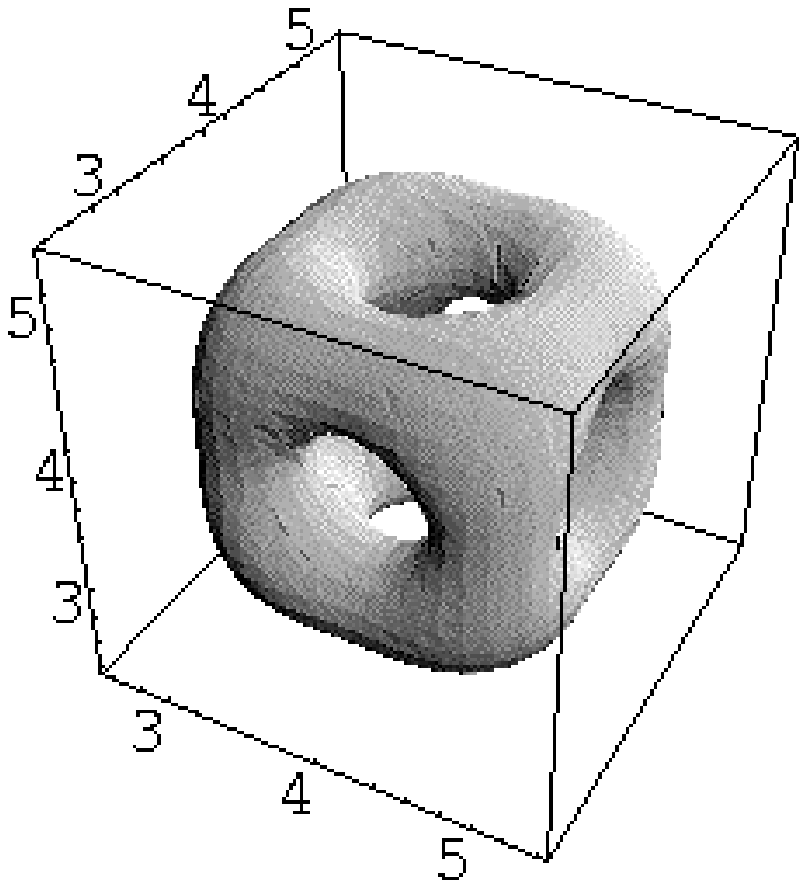}}& &\hbox{\sfig{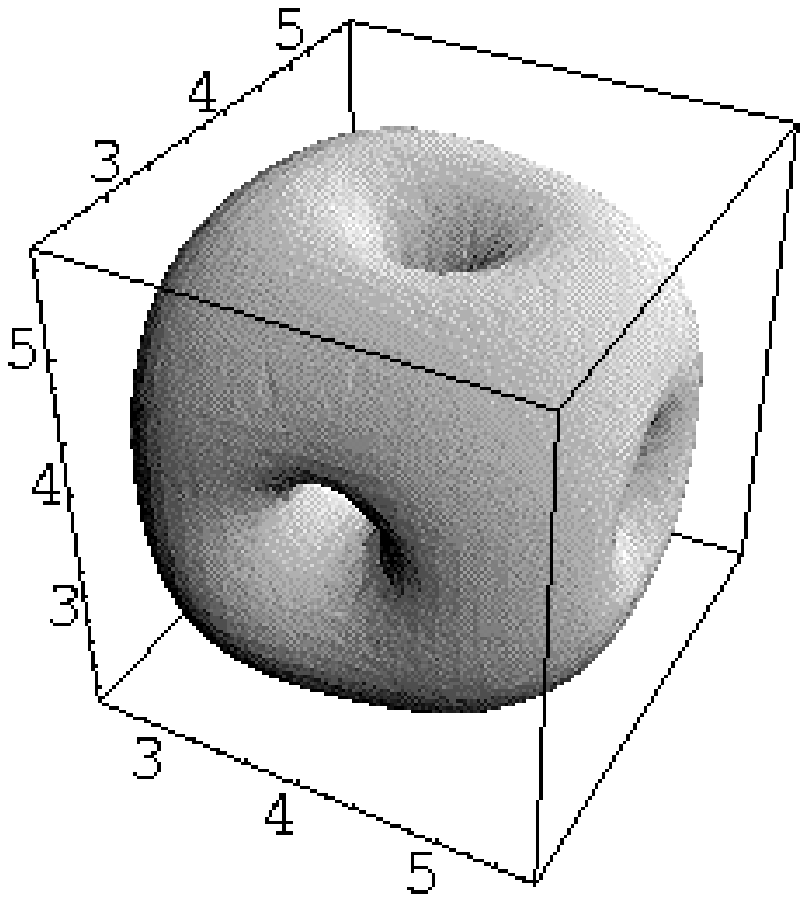}\hfil\sfig{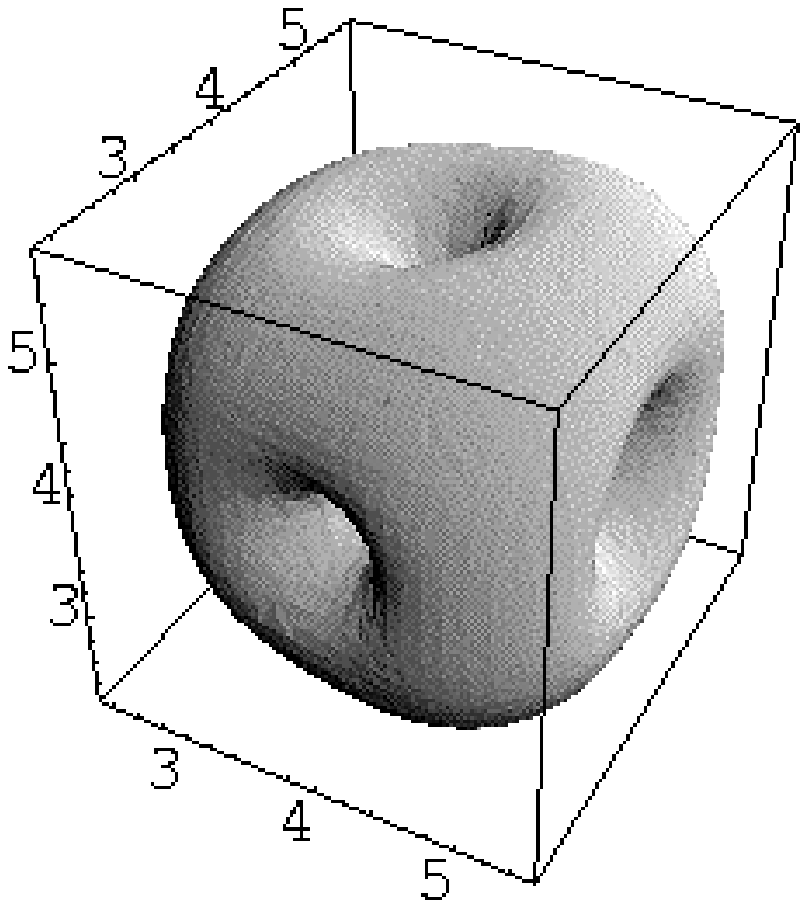}}& &
\hbox{\sfig{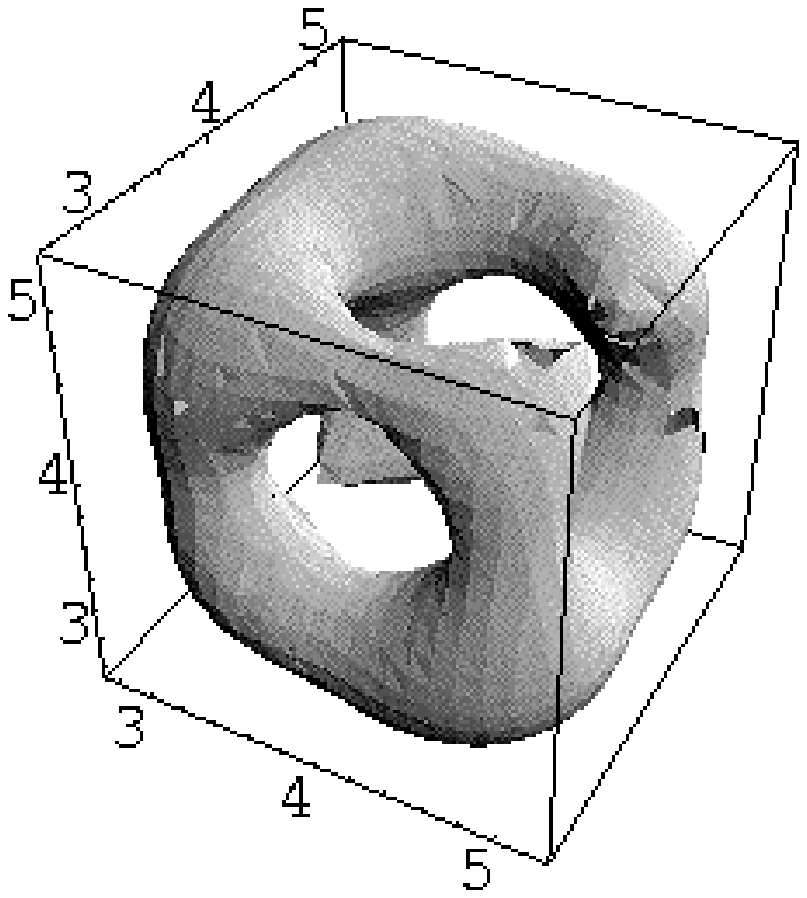}\hfil\sfig{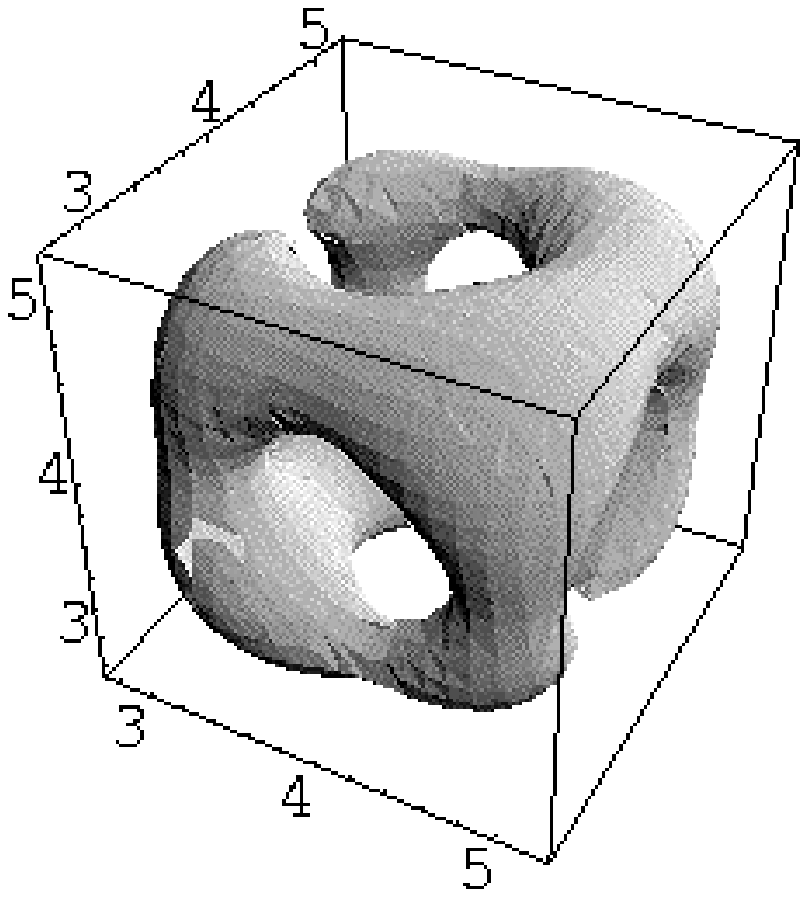}} \\
 & \\   
\end{tabular}
\caption{Contours of constant baryon density for the B=4
soliton, combined with its normal modes, as indexed by their
frequencies in Figure \protect\fig{spectrum}. These modes were studied
in a box of size $8{\times}8{\times}8$, with a grid spacing $\Delta_x
= {1 \over 8}$ in Skyrme units.  For comparison, the soliton itself is
a cube roughly $2{\times}2{\times}2$ in these units. For the case $\omega=.367$ 
two orthogonal modes of the degenerate multiplet are shown,
in other cases a single mode only is shown.}
\labfig{bigplot}
\end{figure*}
\begin{table*}[t]
\begin{tabular}{|lccp{4.5in}|}
 \bf Frequency & \bf Degeneracy & \bf Symmetry & \bf Description \\
\hline\hline
.07 & 3 &$F_1^{+}$& 
Rotations of the soliton.  This is a zero mode broken
by the finite box size. \\ \hline
.367 & 2 &$E^{+}$& Lowest 
 vibrational modes. One mode, $\delta^1$, alternately
pulls the
$B=4$ cube into two $B=2$ donuts in two perpendicular
directions.
The orthogonal mode, $\delta^2$, pulls it into
four $B=1$ edges one way and 
two $B=2$ donuts the other.
\\ \hline
.405 & 1 & $A_2^{-}$  & The corners of the cube make two interlacing 
tetrahedra. This mode pulls one tetrahedron out into four $B=1$ corners, 
pushing
the other one in. \\ \hline
.419 & 3 &$F_2^{+}$ & Deform two opposite faces of the cube into rhombuses.
\\ \hline
.513 & 3 &$F_2^{-}$& Deform the cube by pulling two opposite edges on one 
face, and the two perpendicular edges on the opposite face.
This takes the cube to four $B=1$ edges. 
\\ \hline
.545 & 2 &$E^{+}$ & Two of the pion $k=0$ modes.
 \\ \hline 
.587 & 1 & $A_2^{-}$  & The remaining pion $k=0$ mode, with tetrahedral symmetry. \\ \hline
.605 & 1 & $A_1^{+}$ & The breathing mode, with the full cubic symmetry of the soliton. \\ \hline
.655 & 3 &$F_2^{-}$ & One face of the cube inflates, while the opposite face
deflates. \\ \hline
.738 & 3 &$F_2^{+}$ & One pair of diagonally opposite edges inflates, the parallel pair deflates. \\ \hline
.908 & 3 & & The lowest nonzero ($k=1,0,0$) pion radiative mode.
\end{tabular}
\caption{Description of the Skyrme field $B=4$ normal modes marked in Figure 1.
The notation of Hamermesh \protect\cite{Hamermesh} is used for the
representations of the cubic group $O_h$;
superscripts denote parity.}
\label{modetable}
\end{table*}

\end{document}